\documentclass[12pt]{article} \input epsf.tex
\usepackage{epsfig,graphicx,psfrag}
\usepackage{axodraw}
\usepackage{amsmath,bbm,amsfonts,amssymb}
\usepackage{latexsym}\usepackage{stmaryrd}

\newcommand{\be}{\begin{equation}}
\newcommand{\ee}{\end{equation}}
\newcommand{\bea}{\begin{eqnarray}}
\newcommand{\eea}{\end{eqnarray}}
\newcommand{\bear}{\begin{eqnarray}}
\newcommand{\eear}{\end{eqnarray}}
\newcommand{\ba}{\begin{array}}
\newcommand{\ea}{\end{array}}
\newcommand{\lae}{\begin{array}{c}\,\sim\vspace{-21pt}\\<
\end{array}}
\newcommand{\gae}{\begin{array}{c}\,\sim\vspace{-21pt}\\>
\end{array}}

\newcommand{\beq}{\begin{equation}}
\newcommand{\eeq}{\end{equation}}
\newcommand{\beqs}{\begin{eqnarray}}
\newcommand{\eeqs}{\end{eqnarray}}

\begin{document}

\baselineskip=18pt \pagestyle{plain} \setcounter{page}{1}

\vspace*{-0.8cm}

\noindent \makebox[11.6cm][l]{\small \hspace*{-.2cm} February 4, 2009; Revised April 10, 2009}
{\small Fermilab-Pub-08-571-T}  
\\ [-2mm]

\begin{center}
{\Large \bf Prospects for top-prime quark discovery \\ [3mm] at the Tevatron} \\ [9mm]

{\normalsize \bf Bogdan A. Dobrescu, Kyoungchul Kong, Rakhi Mahbubani \\ [4mm]
{\small {\it
Theoretical Physics Department, Fermilab, Batavia, IL 60510, USA }}\\
}
\end{center}

\vspace*{0.1cm}

\begin{abstract}
We show that a top-prime quark as heavy as 600 GeV can be discovered at
the Tevatron, provided it is resonantly pair-produced via a vector
color octet.  If the top-prime originates from a vectorlike quark, 
then the production of a
single top-prime in association with a top may also be observable,
even through its decay into a Higgs boson and a top.
A color octet  with mass of about 1 TeV, which decays into a top-prime pair,
may account for the CDF excess of semileptonic $(Wj)(Wj)$
events.
\end{abstract}


\section{Introduction} \setcounter{equation}{0}

The top-quark discovery at the Tevatron \cite{Abe:1995hr}, 
made more than a decade ago, 
has completed the quark sector of the standard model. It is imperative to ask
whether physics beyond the standard model includes additional quarks,
and in particular, whether there are any heavier quarks with collider signatures
similar to those of the top quark.

All standard model fermions are chiral ({\it i.e.}, their left- and right-handed 
components have different gauge charges). 
Additional chiral quarks could exist, although a variety 
of nontrivial constraints need to be satisfied \cite{Frampton:1999xi,Kribs:2007nz,Hung:2007ak}. 
Most importantly, electroweak observables are 
highly sensitive to the presence of chiral fermions \cite{Peskin:1991sw}.

A simpler possibility would be the existence of vectorlike ({\it i.e.}, 
non-chiral) fermions. These can easily evade all experimental constraints
because their effects decouple in the limit of large fermion masses.
Given that the masses of vectorlike fermions 
are not protected by the electroweak symmetry, they are likely to be 
larger than the electroweak scale. As a result, even if 
vectorlike fermions exist, it is natural that they have not been 
discovered so far. 

A vectorlike quark that mixes with the top quark may decay into $Wb$
with a sizable branching fraction,
appearing in collider experiments as a heavier copy of the top quark 
\cite{Frampton:1999xi,Choudhury:2001hs,Han:2005ru,:2006jm}. 
For a  region of parameter space, the same is true for an up-type quark belonging to a chiral
4th generation \cite{Hung:2007ak}.
The CDF Collaboration has searched for $t^\prime$ pair
production, where $t^\prime$ is a hypothetical quark assumed to decay 
to a $W$ boson and a jet, with 760 pb$^{-1}$ of data in Run II of the Tevatron, 
setting a lower mass limit of 256 GeV at the 95\% CL \cite{:2008nf}.
More recently, with 2.8 fb$^{-1}$, the CDF limit on the $t^\prime$ mass
has been raised to 311 GeV \cite{Lister:2008is}.
Given the record luminosities achieved recently at the Tevatron, and various
improvements of the search techniques, the
CDF and D0 experiments may become sensitive to $t^\prime$ 
pairs produced via QCD for $t^\prime$ masses up to roughly 350 GeV. 

Here we show that the top-prime production cross section at 
hadron colliders could be substantially higher than the QCD prediction if a ``gluon-prime'', 
{\it i.e.} a massive color-octet vector boson, is present in the theory. 
This provides motivation for the search of significantly heavier $t^\prime$ 
quarks, and in particular allows the interpretation of some excess events 
observed by the CDF Collaboration \cite{Lister:2008is} in terms of a 
$t^\prime$ with mass of about 450 GeV.
Specifically, we consider an extension of the standard model which includes 
a vectorlike fermion that has the same gauge charges as the right-handed top quark,
and a gluon-prime arising from an $SU(3) \times SU(3)$ extension of QCD
similar to those presented in \cite{Hill:1991at} (the gluon-prime in that case 
is referred to as a ``coloron''). 
We find that this model predicts  several interesting 
collider signatures, and propose further ways of testing the origin of the 
CDF excess.


The two hypothetical particles discussed here are part of various 
models for TeV scale physics. In the top-quark seesaw 
model \cite{Dobrescu:1997nm, Chivukula:1998wd, Collins:1999rz},
the vectorlike quark binds to the top-quark forming a Higgs boson,
with binding provided by the coloron.
In certain models with a flat \cite{Cheng:1999bg} or warped 
\cite{Carena:2006bn} extra dimension, 
the vectorlike quark and the gluon-prime appear as Kaluza-Klein modes.
 Rather than imposing here any of the constraints among parameters 
present in these models, we consider a generic renormalizable theory that includes
these two particles. 

In Section 2 we review the properties of the vectorlike quark,
in the absence of a gluon-prime. We then analyze the $t^\prime$+$G_\mu^\prime$ model
in Section 3, and present its implications for Tevatron searches in Section 4.
A concluding discussion is given in Section 5.

\smallskip 
\section{Vectorlike origin of the top-prime quark}
\setcounter{equation}{0}\label{sec:tprime}

Let us start with the standard model plus a single vectorlike quark,
labelled by $\chi$, which transforms as $(3,1, 2/3)$ under 
the $SU(3)_c\times SU(2)_W \times U(1)_Y$ gauge group.
This up-type vectorlike quark, having electric charge +2/3, may mix with the top quark.
The mixings of $\chi$ with the first two generations of quarks are expected to be small, 
and we will neglect them.
The Lagrangian includes two gauge invariant 
quark mass terms and two Yukawa interactions of $\chi$ and $u^3$ to the Higgs doublet,
where $u^3$ is the standard model up-type quark of the third generation in the 
gauge eigenstate basis.
Given that $SU(2)$ transformations that mix $\chi_R$  and $u^3_R$ are not physically 
observable, we can choose one of the Yukawa couplings to vanish.
Thus, after electroweak symmetry breaking, the quark mass matrix and interactions with the 
Higgs boson  are given by 
\begin{equation}
{\cal L} = - \left( \overline{u}^3_L \ , \ \overline{\chi}_L \right)
\left( \ba{cc} \lambda_t \left(v_H + h^0 /\sqrt{2}\right) & 0 \\
    M_0 & M_\chi \ea \right)
\left( \ba{c} u^3_R \\ \chi_R \ea \right) + {\rm H.c.} ~,
\label{massterm}
\end{equation}
where $h^0$ is the Higgs boson and $v_H \simeq 174$ GeV is the VEV of the Higgs doublet.
The two mass parameters, $M_0$ and $M_\chi$, and 
the $\lambda_t$ Yukawa coupling are taken to be real parameters, 
as their complex phases can be absorbed 
by $U(1)$ transformations of the quark fields.

To relate these three real parameters of the Lagrangian to physical observables, we 
transform the gauge eigenstates $u^3_{L,R}$ and $\chi_{L,R}$ to the mass 
eigenstates $t_{L,R}$, $t^\prime_{L,R}$, where
$t$ is the top quark observed at the Tevatron, of mass $m_t\approx 173$ GeV,
and $t^\prime$ is a new quark of mass $m_{t^\prime}$, which remains to be discovered.
The relation between the two bases depends on two angles, $\theta_L$ and $\theta_R$,
\bear
&& \left( \ba{c} t_{L,R} \\ t^\prime_{L,R} \ea \right)
=
\left( \ba{cc} c_{L,R} & -s_{L,R} \\ s_{L,R} & c_{L,R} \ea \right)
\left( \ba{c} u^3_{L,R} \\ \chi_{L,R} \ea \right)  ~,
\eear
where $s_{L,R}$ and $c_{L,R}$ are short-hand notations for $\sin\theta_{L,R}$, 
and $\cos\theta_{L,R}$, respectively. 
As explained above, no physical observable depends on $\theta_R$
(the situation changes in the model presented in the next section). 
The mixing angle $\theta_{L}$ affects the electroweak interactions of the 
top quark as well as the Yukawa couplings of the Higgs boson.
The relations between the physical parameters $m_t, m_t^\prime$, $\theta_{L}$ 
and the initial parameters $\lambda_t, M_\chi, M_0$ are given by
\begin{equation}
m_{t,t^\prime}^2 = \frac{1}{2} \left( M_\chi^2 + M_0^2 +  \lambda_{t}^2v_H^2 \right)
\left[ 1 \mp \sqrt{ 1 - 
\left( \frac{2\lambda_t v_H  M_\chi}{M_\chi^2 + M_0^2 + \lambda_{t}^2 v_H^2} \right)^{\! 2} } 
\; \right]
\label{mtmc}
\end{equation}
for the masses, and by
\begin{equation}
\tan 2\theta_L = \frac{2 M_0 \lambda_{t}v_H}{M_\chi^2+ M_0^2-\lambda_t^2 v_H^2} ~~,
\label{mix}
\end{equation}
for the  $\theta_L$ mixing angle, which is taken to be between 0 and $\pi/2$. 
The $\theta_R$ mixing angle is given in terms of  $m_t, m_{t^\prime}$ and $\theta_{L}$ 
by
\begin{equation}
s_R^2 = \frac{ s_L^2 m_{t^\prime}^2}{s_L^2 m_{t^\prime}^2 + c_L^2 m_t^2 } ~.
\label{mix-R}
\end{equation}

Although $s_L$ and $m_{t^\prime}$ are independent parameters, $s_L$ cannot vary 
over the entire range 0 to 1 when $m_{t'}$ is large enough.
To see this, it is useful to express $s_L$ in terms of $\lambda_t v_H$, 
$m_t$, and $m_{t^\prime}$:
\begin{equation}
s_L=\sqrt{ \frac{ \lambda_t^2 v_H^2 - m_t^2 }{ m_{t'}^2 - m_t^2 } }  ~ .
\end{equation}   
When $\lambda_t \gg 1$ the above relation implies
\begin{equation}
s_L = \frac{ \lambda_t v_H } { m_{t'} } 
\left[ 1 + O \left( \frac{ m_t^2 }{ m_{t'}^2 } \right) 
+ O \left( \frac{ m_t^2 }{ \lambda_t^2 v_H^2 } \right) \right] ~.
\label{upper}
\end{equation}  
The Yukawa coupling $\lambda_t$ is limited by perturbativity, so that Eq.~(\ref{upper})
gives an upper limit on $s_L$. 
For $m_{t^\prime} \rightarrow \infty$, we see that the mixing vanishes ($s_L \rightarrow 0$)
so that the new physics decouples from the standard model. 
For top-prime masses accessible at the Tevatron, however,
the upper limit from (\ref{upper}) can be ignored because it is above 1.

The interactions of $t$ and $t^\prime$ 
with the electroweak bosons depend on $\theta_{L}$ \cite{:2006jm}.
The charged-current interactions are
\begin{equation}
\frac{g}{\sqrt{2}}\, W_\mu^+ \;
\overline{b}_L\gamma_\mu \left( c_L t_L +s_L t^\prime_L \right) + {\rm H.c.}
\label{wtpb}
\end{equation}
where 
$g \equiv e / \sin \theta_W$ is the $SU(2)_W$ gauge coupling.  
Given that the measurement of single-top production at the Tevatron sets a limit on the 
coefficient of the $t$-$W$-$b$ coupling\footnote{The D0 measurement
\cite{Abazov:2009ii} is $|V_{tb}| = 1.07\pm 0.12$, 
and the CDF one \cite{Aaltonen:2009jj} is 
$|V_{tb}| = 0.91\pm 0.11\pm 0.07$. Combining these two measurements in quadrature
gives $|V_{tb}| > 0.82$ at the 95\% CL. We use this result only as a rough estimate for the 
combined limit; a proper combination of measurements, which takes into account correlations, 
would need to be performed by the CDF and D0 Collaborations. Note that the D0 measurement by 
itself gives almost the same limit: $|V_{tb}| > 0.83$ at the 95\% CL.} 
$c_L \simeq V_{tb} \lae 0.82$, 
we find a nontrivial constraint: 
\be
s_L < 0.57 ~.
\label{eq:sL}
\ee

The $Z$ boson has modified interactions with the left-handed quarks, 
including a flavor-changing  $t$-$t^\prime$ current:
\bear
\frac{g}{\cos\theta_W}Z_\mu && \hspace*{-1.5em} \left[
\left( \frac{c_L^2 }{2} - \frac{2}{3} \sin^2\!\theta_W \right) \overline{t}_L\gamma_\mu t_L
+\left( \frac{ s_L^2}{2}  - \frac{2}{3} \sin^2\!\theta_W \right) \overline{t^\prime}_L \gamma_\mu t^\prime_L \right.
\nonumber \\ [0.5em] && \left.
+\; \frac{s_L c_L}{2} \left( \overline{t^\prime}_L \gamma_\mu  t_L  + {\rm H.c.}\right)\right] ~~.
\eear
The interactions of $t_R$ and $t^\prime_R$ with  the $Z$ boson are
identical with those of the right-handed top quark in the standard 
model.
The Higgs interactions with $t$ and $t^\prime$ can be expressed
in terms of $\theta_L$, $m_t/v_H$ and $m_t^\prime/v_H$:
\be
 \frac{-1}{v_H\sqrt{2}} h^0 \left(c_L^2 m_t \; \overline{t}_L t_R 
+ s_L^2 m_{t^\prime} \; \overline{t^\prime}_L t^\prime_R 
+ c_L s_L m_{t^\prime} \; \overline{t}_L t^\prime_R 
+ c_L s_L m_t \; \overline{t^\prime}_L t_R  \right)  +  {\rm H.c.} 
\ee

The modified electroweak couplings of the $t$ quark as well as the 
new couplings of the  $W$ and $Z$ bosons to the $t^\prime$ quark have an
impact on electroweak observables. Most notably, the $W$ and $Z$ masses
get one-loop corrections such that the $T$ parameter, which measures weak-isospin
violation, is given by \cite{Chivukula:1998wd}
\be 
 T = \frac{3  }{16 \pi  \sin^2\!\theta_W }  
\frac{m_t^2 s_L^2 }{M_W^2} \left( s_L^2 \frac{m_{t^\prime}^2}{m_t^2} 
+ \frac{4 c_L^2 m_{t^\prime}^2}{m_{t^\prime}^2 - m_t^2}
\ln \frac{m_{t^\prime}}{m_t} - 1 - c_L^2 \right) - \Delta (M_h) ~ ~.
\label{eq:T-parameter}
\ee
Here $\Delta (M_h) \ge 0$ is the contribution due to Higgs loops, and depends only on 
the Higgs mass. Using the standard model with $M_h = 115$ GeV as the reference point,
the 95\% confidence limit (assuming an optimal contribution to the $S$ parameter)
is $T \lesssim 0.36$ \cite{Kribs:2007nz}, and
$\Delta$ varies from 0 to $0.15$ as $M_h$ varies from 115 to 500 GeV.
Fixing $m_{t^\prime} = 450$ GeV,
we find that the limit on the $t-t^\prime$ mixing  ranges from
$s_L \lesssim 0.32$ for $M_h = 115$ GeV to $s_L \lesssim 0.38$ for $M_h = 500$ GeV.
Although this limit is more stringent than Eq.~(\ref{eq:sL}), it is  less 
robust: new physics may relax the electroweak fit without being discovered
at the Tevatron or LHC. For example, leptophobic $Z^\prime$ bosons or 
complex Higgs triplets can give negative contributions to $T$, allowing larger values 
for $s_L$. For this reason we will not use
the electroweak constraints in what follows.

The charged-current interactions induce the $t^\prime \rightarrow W^+ b$ decay,
while the flavor-changing neutral-current interactions  
induce the $t^\prime \rightarrow Z^0 t$ decay, assuming 
$m_{t^\prime} \gtrsim 264$ GeV.
The tree-level decay widths are given by
\begin{eqnarray} \label{eq:widths}
&& \hspace*{-.5em} \Gamma(t^\prime \rightarrow W^+ b) = \frac{s_L^2 m_{t^\prime}^3}{32 \pi v_H^2} 
\left(1 - \frac{M_W^2}{m_{t^\prime}^2} \right)^{\! 2} \left(1 + \frac{2M_W^2}{m_{t^\prime}^2} \right) ~,
\nonumber \\ [3mm] && \hspace*{-.5em} 
\Gamma(t^\prime \rightarrow Z^0 t ) = \frac{c_L^2 s_L^2 m_{t^\prime}^3}{64 \pi v_H^2} 
\left[\left(1 - \frac{m_t^2}{m_{t^\prime}^2}\right)^{\! 3}
+ O\left(\frac{M_Z^4}{m_{t^\prime}^4}\right)\right] ~.
\end{eqnarray}
The Higgs interactions allow the $t^\prime \rightarrow h^0 t$ decay, provided 
$m_{t^\prime} > M_h + m_t$:
\be
\label{eq:higgs-width}
\Gamma(t^\prime \rightarrow h^0 t ) = \frac{c_L^2 s_L^2 m_{t^\prime}^3}{64 \pi v_H^2} 
\left(1\! + \!\frac{6m_t^2\! -\! M_h^2}{m_{t^\prime}^2}  
+\! \frac{m_t^4 - m_t^2 M_h^2}{m_{t^\prime}^4} \right)
\beta\!\left(\frac{m_t^2}{m_{t^\prime}^2},\frac{M_h^2}{m_{t^\prime}^2} \right)
~~,
\ee
where $\beta$ is the relative velocity of the decay products:
\be
\beta\!\left(x_1,x_2\right) = \left[ \left(1-x_1-x_2\right)^2-4x_1x_2 \right]^{1/2} ~~.
\label{eq:beta}
\ee
Eqs.~(\ref{eq:widths}) and (\ref{eq:higgs-width}) agree with the results given in 
\cite{AguilarSaavedra:2005pv}.

\begin{figure}[t]
\centerline{
\epsfig{file=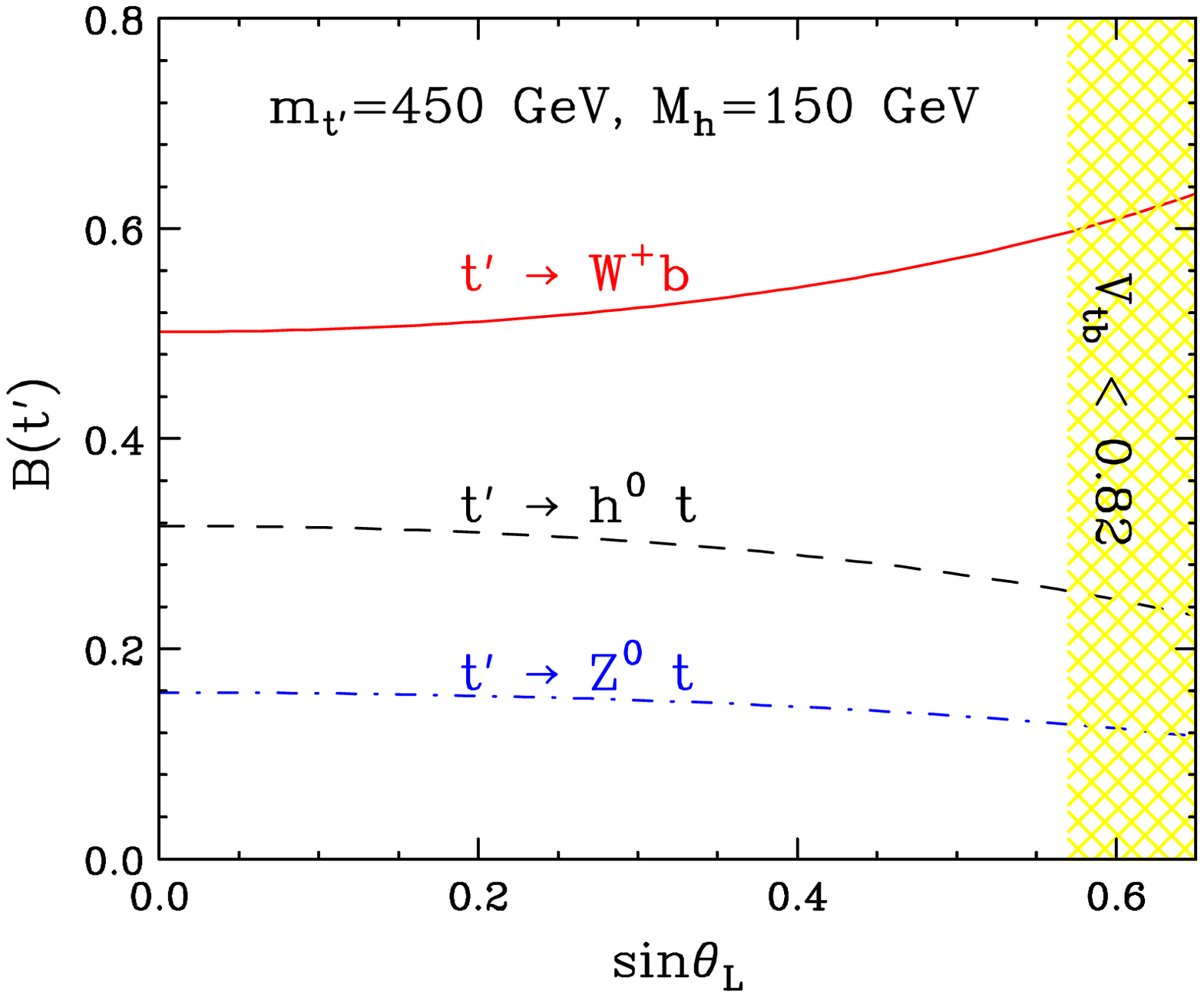,  width=7.95cm,angle=0} \hspace*{0.1cm}
\epsfig{file=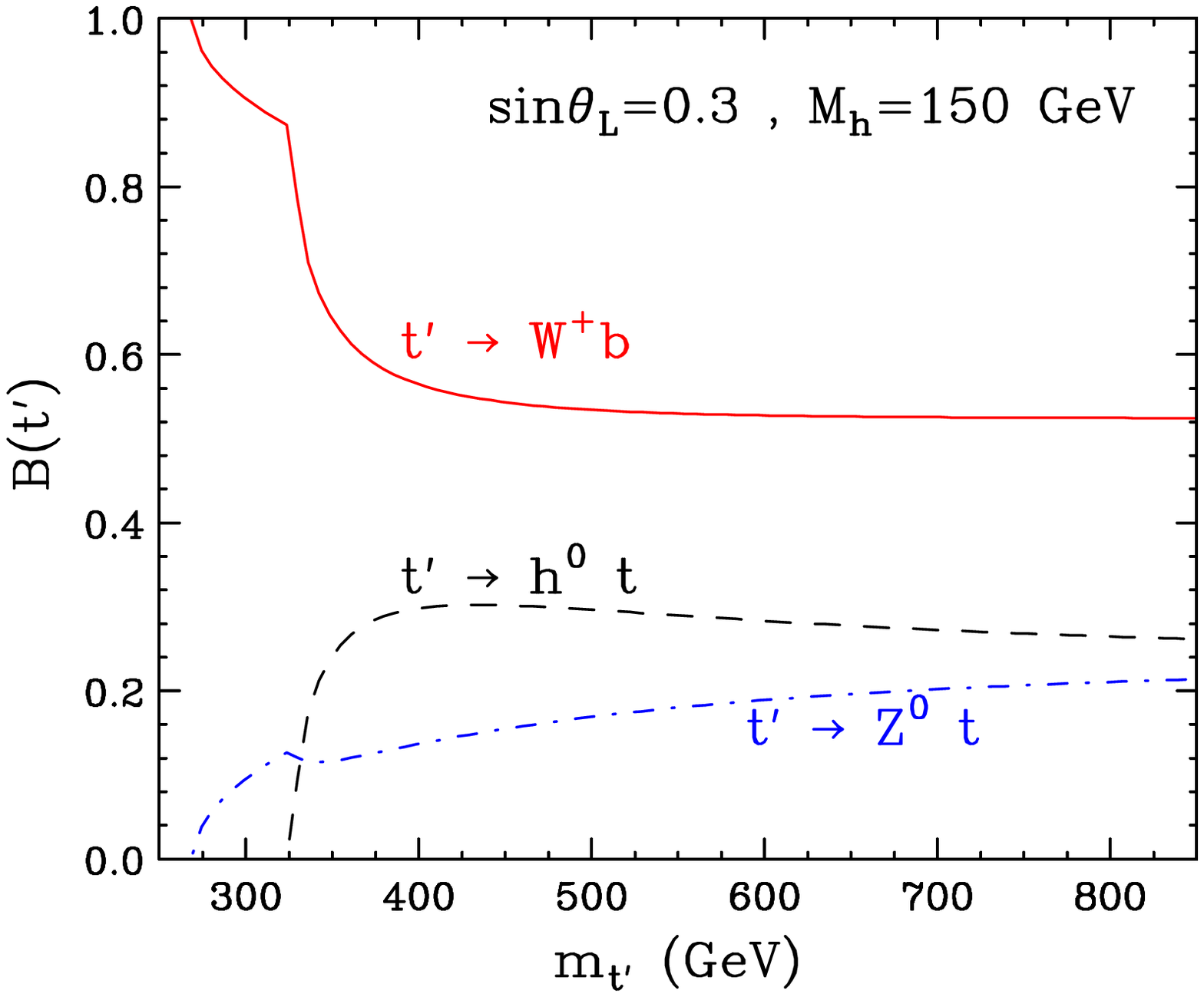, width=7.8cm,angle=0} }
\caption{Branching fractions of $t^\prime$ for a Higgs mass $M_h = 150$ GeV, 
as a function of either $s_L\equiv \sin\theta_L$ or $m_{t^\prime}$.
The shaded region on the first plot is excluded by Eq.~(\ref{eq:sL}).
}
\label{fig:brs}
\end{figure} 

In the limit of $m_{t^\prime} \gg m_t+ M_h$  the branching fractions for the 
$t^\prime$ decays are
\begin{eqnarray} 
&& B(t^\prime \rightarrow W^+ b) = \frac{1}{1 + c_L^2} \ge 50\% ~,
\nonumber \\ [2mm]
&& B(t^\prime \rightarrow Z^0 t ) = B(t^\prime \rightarrow h^0 t ) = \frac{c_L^2}{2\left(1 + c_L^2\right)} \le 25\% ~.
\end{eqnarray}
Given that the $t^\prime$ quark compensates for 
a heavier Higgs boson in the electroweak fits (see Eq.~(\ref{eq:T-parameter})
and Refs.~\cite{Collins:1999rz, Chivukula:2000px}), allowing even 
$M_h \sim 500$ GeV, the $t^\prime \rightarrow h^0 t$ decay could be strongly phase-space 
suppressed for $m_{t^\prime}$ as large as $\sim 700$ GeV.
If $m_{t^\prime} < m_t+ M_h$, then the $t^\prime \rightarrow W^+ b$ branching fraction 
becomes even larger: $B(t^\prime \rightarrow W^+ b) > 2/(2 + c_L^2) > 2/3$, where the first 
inequality is due to the phase-space suppression of the $t^\prime \to Z^0 t$ decay.
As an example, the branching fractions of $t'$ for $M_h=150$ GeV are shown in Figure \ref{fig:brs} 
as a function of $s_L$ for fixed $m_{t^\prime}$, and as a function of  $m_{t^\prime}$
for fixed $s_L$.

It is interesting  that the top-prime lifetime can be long in the
$s_L\to 0$  limit \cite{Frampton:1999xi}.  
The decay length for a top-prime moving at the speed $\beta_{t'}$ 
is given by:
\begin{equation}
L=3 \, {\rm cm}\, \left(\frac{10^{-8}}{s_L}\right)^{\! 2}
\left(\frac{450 \; {\rm GeV}}{m_{t'}}\right)^{\! 3} \;\beta_{t^\prime} ~.
\end{equation}
Therefore the top-prime decay is prompt for $s_L\gtrsim 10^{-7}$.  Nevertheless
the limit of very small $s_L$
is natural in a theory where a symmetry imposes 
$M_0=0$, motivating searches for displaced top-prime vertices.

\bigskip 
\section{A model with gluon-prime and top-prime}
\setcounter{equation}{0}\label{sec:model}

We now extend the model of Section 2 by including a $SU(3)_1 \times SU(3)_2$ gauge symmetry which is spontaneously 
broken down to the diagonal group $SU(3)_c$, identified with 
the gauge symmetry of QCD. This symmetry breaking pattern is due to
the vacuum expectation value 
of a field $\Sigma$ transforming as a bilinear under the two $SU(3)$ groups.
$\Sigma$ may be an elementary scalar so that the theory presented here 
is renormalizable and simple, but the radial degrees of freedom within $\Sigma$
do not play a role in what follows.
The quarks transform under the extended gauge group as shown in Table 1.

\begin{table}[b]
\centering
\renewcommand{\arraystretch}{1.7}
\begin{tabular}{|c||c|c|c|c|}
\hline
                & $SU(3)_1$          & $SU(3)_2$     & $SU(2)_W$   &  $U(1)_Y$  \\
\hline
\hline
SM quarks: \ $q_L^i$, $u_R^i$, $d_R^i$ &   3                &   1   & 2, \ 1, \ 1  & +1/6, \ +2/3, \ $-1/3$ \\
vectorlike quark:  \ $\chi_L,\;\chi_R$     &   1                &   3   & 1 & +2/3 \\
scalar with VEV: \ $\Sigma$        &   3                &$\overline{3}$  & 1 & 0  \\
\hline 
\end{tabular}
\medskip
\caption{Fields charged under the $SU(3)_1 \times SU(3)_2$ gauge group, and their 
electroweak charges. The $i$ upper index of the standard model quarks labels the three generations.}
\end{table}

When $\Sigma$ gets a VEV proportional to the $3\times 3$ unit matrix,
the two  $SU(3)$ gauge bosons $G_\mu^1$ and $G_\mu^2$ mix \cite{Hill:1991at}. 
One linear combination of the two  
$SU(3)$ gauge bosons is the massless gluon of QCD, $G_\mu$,
while the orthogonal  combination $G_\mu'$ is a color-octet boson of spin 1 
and mass $M_G$:
\begin{equation}
\left(
\begin{tabular}{c}
  $G_\mu^1$ \\ [0.2em]
  $G_\mu^2$
\end{tabular}
\right) = \frac{1}{\sqrt{h_1^2+h_2^2}}
\left(
\begin{tabular}{c c}
               $h_2$        &   $- h_1$       \\  [0.2em]
               $h_1$        &    $h_2$
\end{tabular}
\right)  \left(
\begin{tabular}{c}
$G_\mu$\\  [0.2em]
$G'_\mu$
\end{tabular}
\right)    ~, \nonumber
\end{equation}
where $h_1$ and $h_2$ are the $SU(3)_1 \times SU(3)_2$ gauge couplings. 

In the mass eigenstate basis for the fermions and gauge bosons, the gluon interactions with all quarks 
are vectorlike and have a strength set by the QCD gauge coupling 
\be
g_s= \frac{h_1 h_2}{\sqrt{h_1^2+h_2^2}} ~.
\label{eq:gstrong}
\ee
The gluon-prime interaction with  light quarks is also vectorlike, but of different strength:
\begin{equation}
g_s r\;  G_\mu'^a \; \overline{q} \gamma^\mu T^a q  ~,
\label{eq:gs}
\end{equation}
where $T^a$ are the $SU(3)_c$ generators and  
\be
 r \equiv \frac{h_1}{h_2} ~.
\ee
If $ r\ll 1$ ($ r\gg 1$), then the gauge coupling $h_2$ ($h_1$) is large and the theory becomes nonperturbative.
Imposing some loose perturbative condition, $h_2$ ($h_1$) $< 4\pi/\sqrt{N_c}$ with $N_c=3$, and using Eq.~(\ref{eq:gstrong}) with 
$\alpha_s \equiv g_s^2/(4\pi) \approx 0.1$,  we derive the range of values for $ r$ where our tree-level 
results can be trusted:
\be
0.15 \lesssim  r \lesssim 6.7 ~~.
\label{eq:f}
\ee

The gluon-prime interactions with the $t$ and $t^\prime$ quarks are chiral, 
and include flavor-diagonal terms,
\be
g_s G_\mu'^a\left[ \overline{t} \gamma^\mu  \left( g_L P_L + g_R P_R \right) T^a t
+ \overline{t^\prime} \gamma^\mu  
\left( g_L^{\prime\prime} P_L + g_R^{\prime\prime} P_R \right) T^a t^\prime   \right] ~~,
\label{eq:tptp}
\ee
as well as flavor-changing terms, 
\be
g_s G_\mu'^a \;  \overline{t} \gamma^\mu  
\left( g_L^{\prime} P_L + g_R^{\prime} P_R \right) T^a t^\prime  
+ {\rm H.c.}  
\label{eq:tpt}
\ee
The left- and right-handed projection operators are given as usual 
by $P_{L,R} = (1\mp\gamma_5)/2$, the couplings of the left-handed quarks are 
\bear 
g_L = r c^2_L-\frac{s^2_L}{r} \;\;\;\;\; , \;\;\;\;\;
g_L^{\prime\prime} = r s^2_L-\frac{c^2_L }{r}  \;\;\;\;\; , \;\;\;\;\;
g_L^{\prime} = \left(r+\frac{1}{r}\right) \, s_L c_L ~~,
\label{eq:LHcouplings}
\eear
while the right-handed couplings, 
$g_R$, $g_R^{\prime\prime}$ and $g_R^\prime$, are analogous to the left-handed ones
except for the replacements $s_L\to s_R$ and $c_L \to c_R$.

Altogether, the model discussed here has 5 free parameters: 
the masses $m_{t^\prime}$ and $M_G$ of the top-prime and gluon-prime,
the mixing parameter $s_L$ from the top sector, 
the ratio $ r$ of the $SU(3)_1$ and $SU(3)_2$ gauge couplings,
and the standard model Higgs mass $M_h$. 


%
\begin{figure}[t]
\centerline{
\epsfig{file=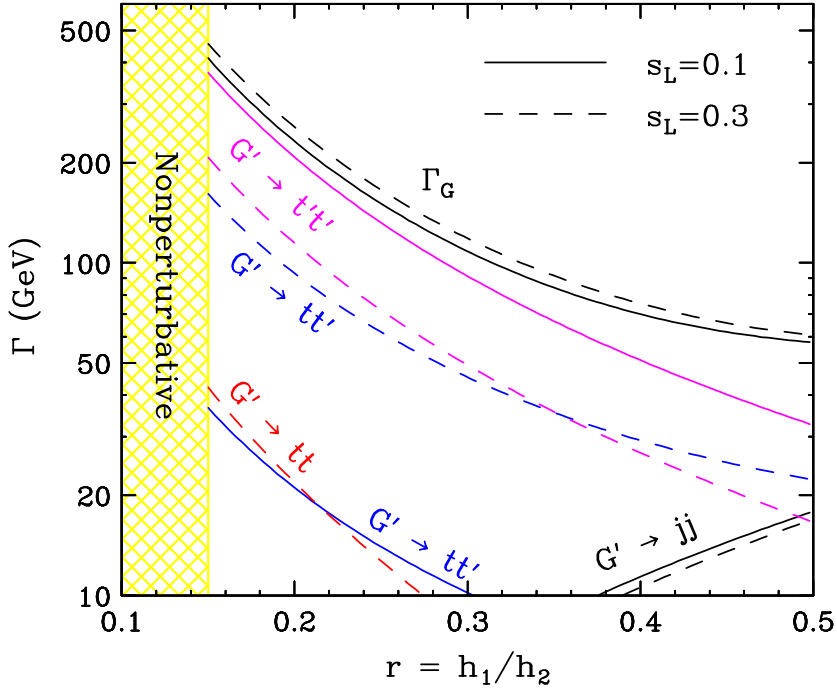,  width=11cm,angle=0}}
\caption{Decay widths of $G^\prime_\mu$ as shown in 
Eqs.~(\ref{eq:jj})-(\ref{eq:ttp}) for its four decay modes,
as a function of the ratio $ r$ of gauge couplings, 
for $M_G = 1$ TeV, $m_{t^\prime} = 450$ GeV, and $s_L =0.1$ (solid lines) or 0.3 
(dashed lines). The top two lines represent the total width.}
\label{fig:widths}
\end{figure} 

Given the couplings shown in Eqs.~(\ref{eq:gs}), (\ref{eq:tptp}) and (\ref{eq:tpt}), 
we can compute the decay widths of $G^\prime_\mu$.
Summing over the standard model quarks other than top, we find
\be
\Gamma\left( G^\prime_\mu \to \sum 
q\bar{q}\right)
= \frac{5 }{6} \, \alpha_s r^2 M_G  ~.
\label{eq:jj}
\ee
The decay width into $t\bar{t}$, for $M_G \gg m_t$, is given by:
\be
\Gamma\left( G^\prime_\mu \to t\bar{t}\right) =
\frac{\alpha_s}{12} \left(g^2_L+ g^2_R\right)
M_G ~ .  
\ee
Including the exact phase-space suppression, the decay width 
into $t^\prime \bar{t^\prime}$ is
\be
\Gamma\left( G^\prime_\mu \to t^\prime\bar{t^\prime} \right) =
\frac{\alpha_s }{12} 
\left[ \left( g^{\prime\prime \, 2}_L + g^{\prime\prime \, 2}_R \right)
\left(1 - \frac{m_{t^\prime}^2}{M_G^2} \right) 
+ 6 g_L^{\prime\prime} g_R^{\prime\prime} \frac{m_{t^\prime}^2}{M_G^2} \right] 
 \left(1 - \frac{4m_{t^\prime}^2}{M_G^2} \right)^{\! 1/2}\!\!\!M_G  ~.
\ee
Finally, the gluon-prime may decay into a top and a top-prime quark, with a width
\be
\Gamma\left( G^\prime_\mu \to t\bar{t^\prime}+t^\prime \bar{t}\right) =
\frac{\alpha_s}{6} 
\left[ \left( g^{\prime \, 2}_L + g^{\prime \, 2}_R \right) 
F\!\left(\frac{m_{t^\prime}^2}{M_G^2},\frac{m_t^2}{M_G^2} \right)
+ 6 g_L^{\prime} g_R^{\prime} \frac{m_{t^\prime} m_t}{M_G^2} \right] 
\beta\!\left(\frac{m_{t^\prime}^2}{M_G^2},\frac{m_t^2}{M_G^2} \right)M_G ~,
\label{eq:ttp}
\ee
where $\beta(x_1,x_2)$ is given in Eq.~(\ref{eq:beta}) and 
\be
F\!\left(x_1,x_2\right)=1-\frac{1}{2}\left(x_1+x_2\right)-\frac{1}{2}\left(x_1-x_2\right)^2 ~.
\label{eq:F}
\ee
In Figure \ref{fig:widths} we plot the tree-level partial widths of the gluon-prime as a function 
of $ r$ for fixed $s_L$.

The total width of $G^\prime_\mu$, $\Gamma_G$, is the sum of the above 
four decay widths. For small $ r$ the total width is large, as shown in  
Figure \ref{fig:widths},
but for $ r \gae 0.3$ the gluon-prime becomes a narrow resonance, with
$\Gamma_G$ less than 10\% of $M_G$.
Note that for $M_G \gg 2 m_{t^\prime}$,
the total width is independent of the quark mixing angles:
\be
\Gamma_G = \frac{\alpha_s}{6} \left( 6r^2 + \frac{1}{r^2}\right) M_G  ~.
\ee

\bigskip 
\section{Resonant production of  top-prime quarks}
\setcounter{equation}{0}\label{sec:production}

Let us now compute the cross sections for quark pair production 
at hadron colliders. We are primarily interested in the case where the $t'\bar{t'}$
production via an $s$-channel gluon-prime is substantially larger 
than the QCD production.
A large $G^\prime_\mu$ 
coupling to light quarks would be in conflict with the limits 
on dijet resonances, but the  $t'\bar{t'}$ cross section can still be substantially
enhanced if the $G^\prime_\mu$ width is small.
In that case we can ignore the interference between the gluon-prime and gluon.
The partonic cross section for $t'\bar{t'}$ production 
as a function of the center-of-mass energy of the collision $\hat{s}$ is
\be
\sigma\left(q\bar{q} \to G^\prime_\mu \to t'\bar{t'}\right) 
=
\frac{\sigma_0}{2} \, \left( 1 - \frac{4m_{t^\prime}^2}{\hat{s}} \right)^{\!\! 1/2}
\bigg [
\left( g^{\prime\prime \, 2}_L + g^{\prime\prime \, 2}_R \right)
\left( 1- \frac{m_{t^\prime}^2}{\hat{s}} \right) 
+ 6 g_L^{\prime\prime} g_R^{\prime\prime}\frac{m_{t^\prime}^2}{\hat{s}}
\bigg ]  ~. \; \; 
\label{equ:cstptpb}
\ee
The resonant part of the above cross section is contained in 
\be
\sigma_0 = \frac{8\pi}{27} \alpha_s^2 r^2 
\; \frac{\hat{s}}{\left(\hat{s}-M_G^2\right)^2 + M_G^2 \Gamma_G^2}  ~,
\label{eq:sigma0}
\ee
where $\Gamma_G$, the total width of the gluon-prime, was computed at the end of Section 3.
Similarly, the $t\bar{t}$ production cross section due to an $s$-channel gluon-prime is given by
replacing $m_{t^\prime} \to m_t$, $g_L^{\prime\prime} \rightarrow g_L$ and  $g_R^{\prime\prime} \rightarrow g_R$ 
in the above expression for $\sigma\!\left(q\bar{q} \to G^\prime_\mu \to t'\bar{t'}\right)$. 
The cross section for producing a top and a top-prime is:
\be\label{equ:csttpb}
\sigma\left(q\bar{q} \to G^\prime_\mu \to t\bar{t'}+ t'\bar{t}\right) = \sigma_0
\, \beta\!\left(\frac{m_{t^\prime}^2}{\hat{s}},\frac{m_t^2}{\hat{s}}\right)
\left[\left( g^{\prime \, 2}_L + g^{\prime \, 2}_R \right)
F\!\left(\frac{m_{t^\prime}^2}{\hat{s}},\frac{m_t^2}{\hat{s}} \right) 
+ 6 g_L^{\prime} g_R^{\prime}\frac{m_t m_{t^\prime}}{\hat{s}}\right] ~,
\ee
with the functions $\beta$ and $F$ given in 
Eqs.~(\ref{eq:beta}) and (\ref{eq:F}), respectively.
The dijet cross section, including 5 quark flavors and neglecting the 
$t$-channel exchange, is simply  
\be
\sigma\!\left(q\bar{q} \to G^\prime_\mu \to jj \right)
\simeq 5 r^2 \sigma_0 ~.
\label{eq:dijet}
\ee

\begin{figure}[t]
\centerline{
\epsfig{file=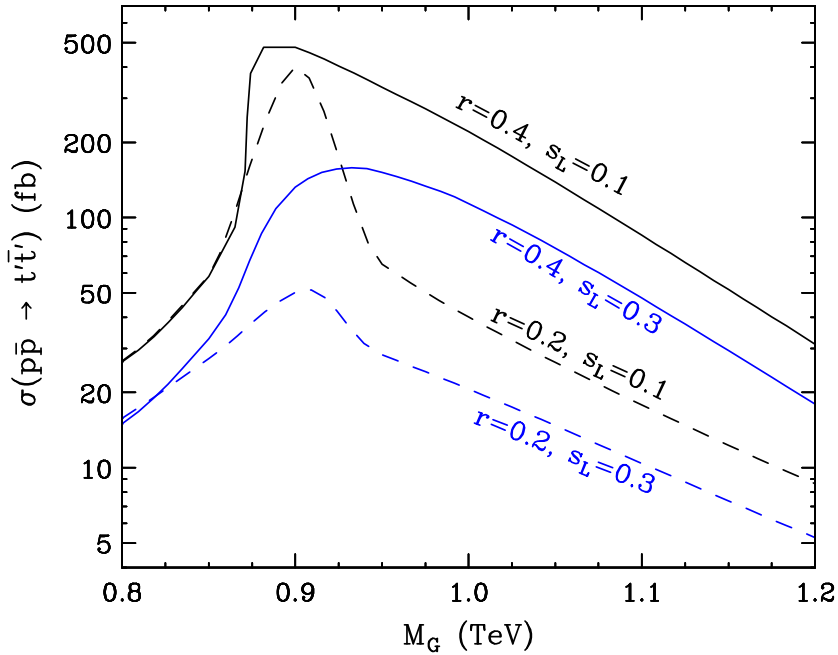, width=11.0cm,angle=0}
}
\caption{Cross section for $t^\prime$ pair production at the Tevatron 
as a function of the $G^\prime_\mu$ mass $M_G$, for $m_{t^\prime} = 450$
GeV, $ r = 0.2$ or 0.4, and $s_L = 0.1$ or 0.3. 
}
\label{fig:cs}
\end{figure}

Convoluting these partonic cross sections with the CTEQ6L \cite{Pumplin:2002vw} parton 
distribution functions, one can obtain the leading-order cross sections at hadron colliders.
We have also implemented the interactions shown in Eqs.~(\ref{eq:gs}), (\ref{eq:tptp}) 
and (\ref{eq:tpt}) in CalcHEP \cite{Pukhov:2004ca} and MadGraph \cite{Maltoni:2002qb}, 
allowing us to more efficiently compute the hadronic cross sections \cite{Tprime}. 
The $p\bar{p}\to G^\prime \to t^\prime \bar{t}^\prime$
cross section in Run II of the Tevatron is shown in Figure~\ref{fig:cs}, 
for a renormalization and factorization scale of $\sqrt{\hat{s}}$.
The threshold behavior close to $M_G \sim 2 m_{t^\prime}$ results in regions 
where a given cross section is consistent with two different $G_\mu^\prime$ masses.
%
%
We do not include any next-to-leading order corrections because they 
have not been computed for this process.

%

\subsection{The $t^\prime$ search in the $(Wj)(Wj)$ final state at the Tevatron}

We now have the tools needed to discuss the existing Tevatron $t^\prime$ searches.
The CDF Collaboration  \cite{Lister:2008is}
has searched for $t^\prime$ pair production in the $\ell + \nu +4j$ channel,
where the lepton and neutrino reconstruct a $W$. 
They then imposed  that two of the jets reconstruct 
a second $W$, and paired the remaining jets with the hadronic and leptonic $W$s
in such a way as to minimize the difference between the invariant mass of the 
jet+$W$ systems.
The latter constraint determines $m_{t^\prime}$. 
In 2.8 fb$^{-1}$ of data they found 7 candidate $t^\prime\bar{t^\prime}$ 
events with reconstructed $m_{t^\prime}$ between 375 and 500 GeV.
The standard model predicts a background of 2.1 events.
Although the excess events may be due to a fluctuation of the background, it is intriguing 
that they can be interpreted as due to the decays of a $t^\prime$ pair with  
$m_{t^\prime} \approx 450$ GeV. Taking into account the resolution, any value of 
$m_{t^\prime}$ between about 400 and 500 GeV would probably fit the data.

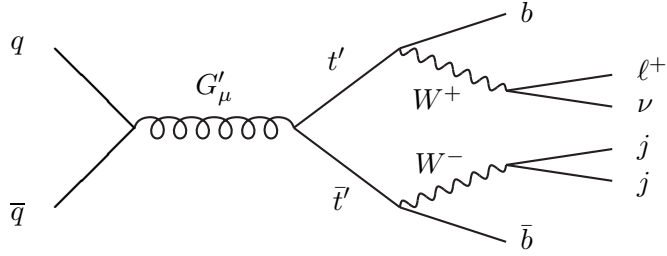
\begin{figure}
\unitlength=1. pt
\SetScale{1.0}
\SetWidth{0.8}      
\begin{center}
\begin{picture}(100,95)(20,-42)
\thicklines
\put(0,0){\line(-1,1){30}}
\put(0,0){\line(-1,-1){30}}
\Gluon(0,0)(60,0){4}{6} 
\Line(60,0)(100,30)
\Line(100,30)(140,43)
\Photon(100,30)(140,14){-2}{6}
\Line(140,14)(180,20)
\Line(140,14)(180,8)
\Line(60,0)(100,-30)
\Line(100,-30)(140,-43)
\Photon(100,-30)(140,-14){2}{6}
\Line(140,-14)(180,-20)
\Line(140,-14)(180,-8)
\put(-47,30){\small $q$}
\put(-47,-35){\small $\overline{q}$}
\put(23,13){\small $G_\mu^\prime$}
\put(73,22){\small $t^\prime$}
\put(75,-31){\small $\bar{t}^\prime$}
\put(146,40){\small $b$}
\put(105,7){\small $W^+$}
\put(106,-17){\small $W^-$}
\put(146,-46){\small $\bar{b}$}
\put(190,18){\small $\ell^+$}
\put(190,6){\small $\nu$}
\put(190,-24){\small $j$}
\put(190,-10){\small $j$}
\end{picture}
\end{center}
%
\caption{Resonant  $t^\prime \bar{t^\prime}$ production followed by 
$t^\prime \to Wb$ decays, with one $W$ decaying leptonically and the other $W$ decaying to jets.}
\label{fig:main-diagram}
\end{figure}

However, the cross section that can account for the observed events is at least an order of magnitude
larger than the QCD production of a $t^\prime$ pair at that mass~\cite{Lister:2008is}.
We propose that this discrepancy is due to the 
resonant enhancement of the cross section when the 
gluon-prime is produced in the $s$-channel, as shown in Figure~\ref{fig:main-diagram}.

One worry about the  $s$-channel resonance interpretation is that, 
in order to be produced at the Tevatron, the new particle must couple to quarks or gluons
and hence would give rise to a dijet resonance that would already have been observed. 
The CDF search for dijet resonances in 1.1 fb$^{-1}$ of data has ruled out a 
gluon-prime of mass up to 1.25 TeV \cite{Aaltonen:2008dn}, assuming that it couples to all quarks
as the usual gluon. In our model, though, the  branching fraction into jets is 
parametrically smaller by a factor $ r^4$: one factor of $ r^2$ coming from the width to dijets,
and another coming from the total width (dominated by the width to  
$t^\prime$ pairs) which  is proportional to $1/ r^2$ for $ r\ll 1$. Furthermore, the production of the gluon-prime is 
suppressed by $ r^2$, due to the smaller coupling to standard model quarks.
As a result, $\sigma(p\bar{p}\to G^\prime \to jj)$ 
decreases as $ r^6$, and the limit of 120 fb set by dijet resonance searches can be 
satisfied for a nontrivial range of parameters:
for $M_G = 1$ TeV, we find $ r \lae 0.4$. Note that the dijet cross section,
given in Eq.~(\ref{eq:dijet}), is only weakly dependent on $s_L$ through $\Gamma_G$.

Searches for $t\bar{t}$ resonances also impose a constraint on our model. 
The partonic cross section, ignoring terms of order $m_t^2/\hat{s}$, may be 
written as
\be\label{equ:csttb}
\sigma\left(q\bar{q} \to G^\prime_\mu \to t\bar{t}\right) 
\approx
\frac{\sigma_0}{2r^2} \, 
\left[   \left( s_L^2 (1+r^2) - r^2 \right)^2 + \left( 
\frac{ s_L^2 m_{t^\prime}^2 (1+r^2)}{m_t^2 + s_L^2 (m_{t^\prime}^2 - m_t^2) }
 - r^2 \right)^{\! 2} \, \right]  ~, \; \; 
\ee
where we used the expression for $s_R$ given in Eq.~(\ref{mix-R}).
The above square bracket is maximized when $s_L \to 1$ as long as $ r < 1$, while for
$ r$ too small $\Gamma_G$ is large so that $\sigma_0$ is small.
Hence, $t\bar{t}$ resonance searches are sensitive to the region of 
large $s_L$ and large $ r$. The current best limit on the cross section of
a $t\bar{t}$ resonance is 570 fb for a resonance of 900 GeV \cite{:2007dia}.

\begin{figure}[t]
\centerline{
\epsfig{file=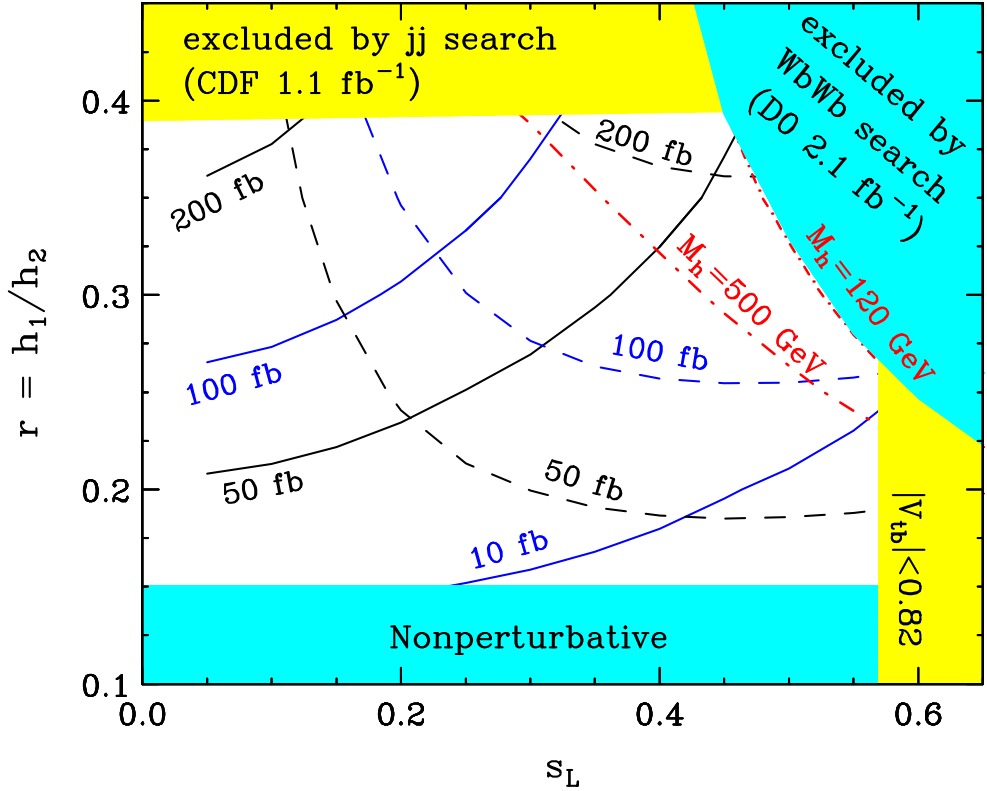,  width=13cm,angle=0} }
\caption{Production cross sections for $t^\prime \bar{t}^\prime$ (solid lines) and 
$t \bar{t}^\prime + t^\prime \bar{t}$ (dashed lines) at the Tevatron, for $M_G= 1$ TeV and  
$m_{t^\prime} = 450$ GeV.
The shaded regions for $s_L > 0.57$ and $ r < 0.15$ are excluded by Eqs.~(\ref{eq:sL})
and  (\ref{eq:f}). 
The shaded region from the upper side of the plot is excluded by the 
search for dijet resonances \cite{Aaltonen:2008dn}.
The search for $(Wb)(Wb)$ resonances \cite{D0-5600} excludes the upper-right corner:
the shaded region for $M_h = 120$ GeV, and everything above the dot-dashed line 
for $M_h = 500$ GeV.
}
\label{fig:r_vs_sL}
\end{figure}

A similar analysis applied to Eq.~(\ref{equ:cstptpb}) shows that
$\sigma\left(q\bar{q} \to G^\prime_\mu \to t'\bar{t'}\right)$
is maximized for $s_L\to 0$.
In this limit, the product of the $G'_\mu$ couplings to light quarks and 
to $t^\prime$ becomes independent of $ r$, but for small $ r$
the $t'\bar{t^\prime}$ cross section still decreases when $ r$ decreases 
because the $G^\prime_\mu$ becomes broader. 
We display the cross section for $t'$ pair production at the Tevatron 
in Figure~\ref{fig:r_vs_sL} as contours  (solid lines) 
in the $s_L$-$r$ plane.  Given that the $t^\prime \to Wb$ branching fraction is approximately $70$\% for large Higgs mass (see Section 2), 
the cross section for $(W^+b)(W^-\bar{b})$ is around half the $t'\bar{t'}$ cross section.

This should be compared with the observed $(Wj)(Wj)$ 
cross section, which must be around $20$ fb in order to give rise to 
5 excess events for an 8.8\% acceptance~\cite{private}, corresponding 
to a $t'\bar{t'}$ cross section of $40$ fb.  Thus, values $ r\gtrsim 0.2$ 
are preferred within the allowed (unshaded) region in Figure~\ref{fig:r_vs_sL}.
With this restriction, the $G'$ width is less than 20\% of its mass (see Figure~\ref{fig:widths}), 
justifying our neglect of interference with the gluon in 
computing the $t'\bar{t'}$ and $t\bar{t}$ cross sections\footnote{The $t'\bar{t'}$ contours in 
the lower part of the allowed region ($ r\lesssim$ 0.2) in Figure~\ref{fig:r_vs_sL} 
are only approximate, since the width of the $G'_\mu$ is large here. 
The interference  with the gluon may easily be included using CalcHEP or MadGraph,
but may complicate the comparison to the data because
experimental plots often assume a narrow width and no interference.}. 

The top-prime production in association with a top quark is also of interest.
Its partonic cross section, given in Eq.~(\ref{equ:csttpb}), vanishes in the $s_L\to 0$ limit,
while for $m_{t^\prime}^2 \gg m_t^2$ it is proportional to 
$s_L^2c_L^2$. Thus, the $t \bar{t}^\prime + t^\prime \bar{t}$ 
cross section reaches its maximum for $s_L$ close to 1/2, away from the maxima of both the
$t^\prime\bar{t^\prime}$ and $t\bar{t}$ cross sections.
In Figure~\ref{fig:r_vs_sL} we show the $t \bar{t}^\prime + t^\prime \bar{t}$ 
cross section contours  (dashed lines) in the $s_L$-$r$ plane.
Strikingly, in the region where the CDF excess events may be explained
and $0.3 \lesssim s_L \lesssim 0.5$ and $0.3 \lesssim  r \lesssim 0.4$, the $(W^+b)(W^-\bar{b})$
cross section from associated $t^\prime$ production
may be up to 5 times larger than that from $t^\prime$ pair production.

It is important to note that there are searches for `$t\bar{t}$ resonances'
from both the CDF and D0 Collaborations that
do not attempt to reconstruct the top mass. 
These are in fact searches for $WbWb$ resonances, and in the context of our model, 
they are sensitive to the {\it sum} of the $t\bar{t}$, $t^\prime\bar{t^\prime}$,
$t^\prime\bar{t}$ and $t\bar{t^\prime}$ resonances.
The CDF search of this type \cite{:2007dz}, with 680 pb$^{-1}$ of data, yielded 
13 events with $WbWb$ invariant masses in the 750 - 1000 GeV range, while the estimated
background is approximately 5 events.
The D0 search \cite{D0-5600}, with 2.1 fb$^{-1}$, set a 
95\% confidence level  upper limit of 210 fb on a $WbWb$ resonance of 1 TeV.
This limit excludes a region in the $s_L$-$r$ plane, which depends on the Higgs mass
because of the $t^\prime \to Wb$ branching fraction. For $M_h = 120$ GeV the  
shaded region in the upper right corner of Figure~\ref{fig:r_vs_sL} is excluded;
for larger Higgs masses the excluded region grows, reaching the dot-dashed line for $M_h = 500$ GeV.


Figure \ref{fig:r_vs_sL} shows that a sizable region of parameter space is 
consistent with the $(Wj)(Wj)$ events observed at CDF when $M_G = 1$ TeV. 
Larger values of $M_G$, up to about 1.2 TeV, can also give a $t'\bar{t^\prime}$ cross 
section above 40 fb, especially for small $s_L$.
Values of $M_G$ below 1 TeV are also fine, but the constraint on $ r$ from dijet resonance 
searches becomes more stringent.
For $M_G$ as low as 850 GeV, the $t'\bar{t^\prime}$ cross section remains above 40 fb,
with the small $ r$ and small $s_L$ region satisfying the dijet constraint (see the 
$r=0.2$ and $s_L = 0.1$ line on Figure~\ref{fig:cs}).

Should the potential $t^\prime$ pair signal at CDF grow with luminosity, and also 
be confirmed by the D0 Collaboration,
then its large cross section will point to an $s$-channel resonance. If this is the case, 
the invariant mass distribution of the full 
$(Wj)(Wj)$ final state should have a peak, representing the observation 
of a new particle (the gluon-prime in our model) as well as a measurement of its mass ($M_G$).
Even if the current excess turns out to be only a fluctuation of the background,
searching for a peak in the distribution of the total invariant mass in each event may 
uncover a top-prime signal that is otherwise swamped by background.
It is thus important for the CDF and D0 Collaborations to impose a mass constraint
on the $(Wj)(Wj)$ sample 
where one $W$ and one jet have a mass peak at $m_{t^\prime}$ and the 
other $W$ and the remaining jet have a mass peak at either $m_t$ or $m_{t^\prime}$.

In addition, detailed studies of the $t'$ can be carried out. 
Its spin and couplings to standard model particles may be measured
by adapting current Tevatron analyses for the top quark, such as $W$-helicity measurements. 
Moreover, the invariant mass distribution of the lepton and $b$-jet from the 
same $t'$ decay, which differs 
from that of the $t$ due to the larger mass of the $t^\prime$,
allows the measurement of the chiral structure of the $t^\prime$-$W$-$b$ coupling and 
$t^\prime$ spin \cite{Burns:2008cp}.
Also, the measurement of the  $t^\prime \bar{t'}$ forward-backward asymmetry will 
probe the chiral structure of the $G^\prime_\mu$ couplings.

\begin{figure}[t]
\centerline{
\epsfig{file=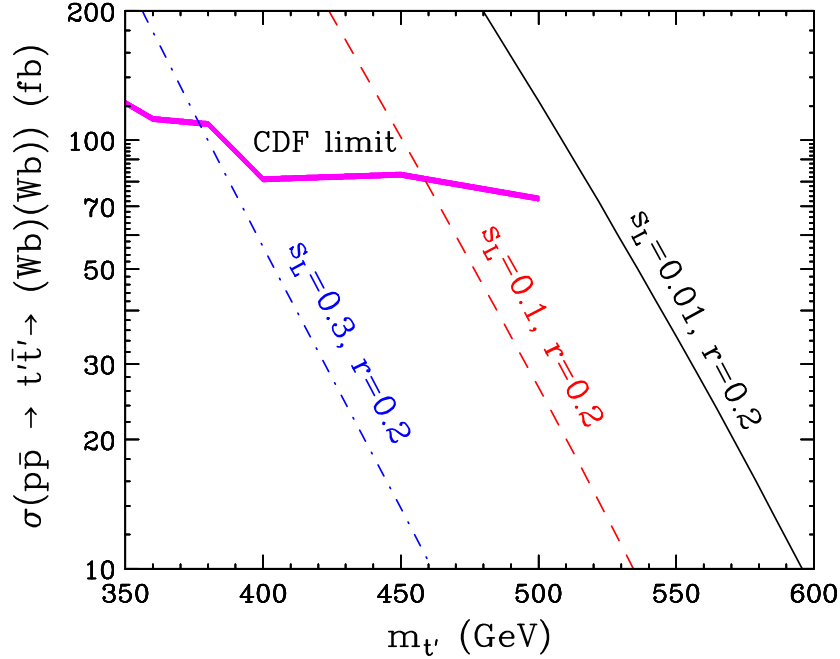, width=11cm,angle=0} 
}
\caption{
Cross section for $p\bar{p} \to G^\prime \to t^\prime \bar{t}^\prime \to (Wb)(Wb)$ at the Tevatron 
as a function of $t^\prime$ mass, 
for $M_G=2m_{t^\prime}$, $M_h = 120$ GeV, $r=0.2$, 
and $s_L=0.01$, $0.1$ or $0.3$.  Also shown is the CDF limit, taken from \cite{Lister:2008is}.
}
\label{fig:cs2}
\end{figure}

Given the large enhancement of $t^\prime\bar{t^\prime}$ production 
in the presence of the $G^\prime_\mu$, it is instructive to estimate the ultimate 
$m_{t^\prime}$ reach of the Tevatron. At the $M_G = 2 m_{t^\prime}$ threshold
the $G^\prime_\mu$ width becomes very small because the $t^\prime\bar{t^\prime}$
decay is no longer kinematically available. For small $s_L$, we get $\Gamma_G \approx 
\alpha_s r^2 M_G$, so that the width is smallest for small $ r$.
As a consequence, at threshold 
the $p\bar{p} \to G^\prime_\mu \to t^\prime\bar{t^\prime}$ cross section
becomes large for small $ r$, in contrast with the small $ r$ region in Figure~\ref{fig:r_vs_sL}. 
Taking $M_G = 1.2$ TeV, $m_{t^\prime} = 600$ GeV, $s_L = 10^{-2}$ and $ r = 0.2$
we find $\sigma(p\bar{p} \to G^\prime_\mu \to t^\prime\bar{t^\prime}) \approx 35$ fb.
For $M_h > m_{t^\prime} - m_t$, the branching fraction into $Wb$ is 
72\%, so that the 
$(W^+b)(W^-\bar{b})$ signal has a cross section  of 18 fb. 
This is huge compared to the QCD pair production, of approximately 
0.1 fb at  $m_{t^\prime} = 600$ GeV.
Assuming a $\sim 9$\% acceptance for a semileptonic $(Wj)(Wj)$ 
event at $m_{t^\prime} = 600$ GeV, 
we find that there will be approximately 10 events in 
10 fb$^{-1}$ of data, while the background is likely to be very small.
We conclude that $m_{t^\prime} = 600$ GeV will be within the reach of the Tevatron
if enough data is analyzed.
For a lighter Higgs boson, the smaller $t^\prime \to Wb$ branching fractions leads to a 
decrease in the $(Wb)(Wb)$ signal.
The $(Wb)(Wb)$ cross section is plotted in Figure \ref{fig:cs2} as a function of 
$t^\prime$ mass for $M_h = 120$ GeV.

%
%

\subsection{Other signatures at hadron colliders}

Also interesting for both $t^\prime$ pair production and 
associated production are the decays $t^\prime \to Z^0 t$ and $t^\prime 
\to h^0 t$.  In particular, the $t' \to h^0 t$ decay may allow the discovery of the
standard model Higgs boson.
Processes with large branching fractions include
\bear
&& p\bar{p} \to G_\mu^\prime \to t^\prime\bar{t} \to 
\; (tZ^0)\, \bar{t} \;\;  / \;\; (t\, h^0) \, \bar{t} \;\; / \;\; (W^+ b)\, \bar{t} \;\; ,
\nonumber \\  [2mm]
&& p\bar{p} \to G_\mu^\prime \to t^\prime\bar{t}^\prime \to 
\; (tZ^0)(W^-\bar{b}) \;\; / \;\; (t\, h^0)(W^-\bar{b}) \;\;  ,
\nonumber 
\eear 
as well as the charge-conjugated processes. The parentheses here indicate that the two particles 
form a mass peak. 
Any of the  $h^0\to b\bar{b}, W^+W^-, Z^0Z^0$ decay modes may be important depending 
on the Higgs boson mass.

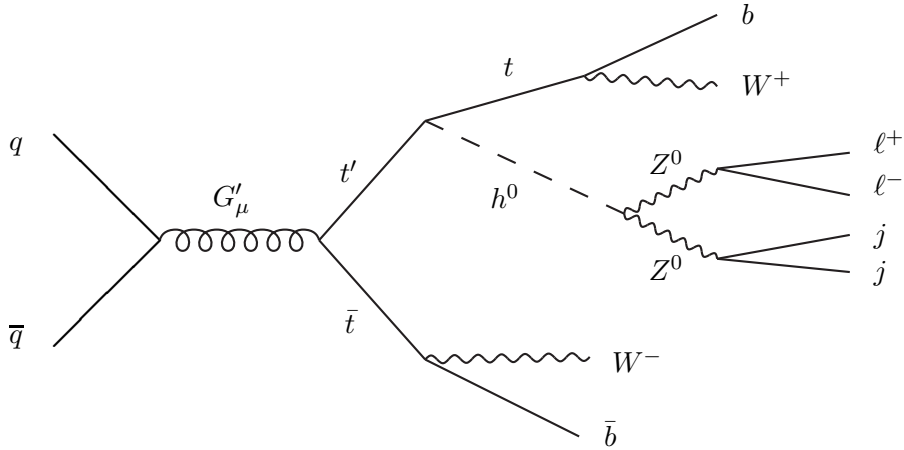
\begin{figure}
\unitlength=1. pt
\SetScale{1.0}
\SetWidth{0.8}      
\begin{center}
\begin{picture}(100,170)(60,-70)
\thicklines
\put(0,0){\line(-1,1){40}}
\put(0,0){\line(-1,-1){40}}
\Gluon(0,0)(60,0){4}{6} 
\Line(60,0)(100,45)
\Line(100,45)(160,62)
\DashLine(100,45)(175,10){7}
\Photon(160,62)(210,58){-1.5}{6}
\Line(160,62)(210,85)
\Photon(175,10)(210,27){1.5}{6}
\Photon(175,10)(210,-7){-1.5}{6}
\Line(210,27)(260,33)
\Line(210,27)(260,17)
\Line(210,-7)(260,2)
\Line(210,-7)(260,-12)
\Line(60,0)(100,-45)
\Line(100,-45)(158,-74)
\Photon(100,-45)(162,-44){1.5}{7}
\put(-57,34){\small $q$}
\put(-57,-39){\small $\overline{q}$}
\put(20,13){\small $G_\mu^\prime$}
\put(68,22){\small $t^\prime$}
\put(130,62){\small $t$}
\put(125,12){\small $h^0$}
\put(70,-35){\small $\bar{t}$}
\put(185,-15){\small $Z^0$}
\put(185,25){\small $Z^0$}
\put(220,82){\small $b$}
\put(220,55){\small $W^+$}
\put(270,33){\small $\ell^+$}
\put(270,17){\small $\ell^-$}
\put(270,-1){\small $j$}
\put(270,-15){\small $j$}
\put(171,-50){\small $W^-$}
\put(168,-78){\small $\bar{b}$}
\end{picture}
\end{center}
%
\caption{Resonant  $t^\prime \bar{t}$ production followed by 
$t^\prime \to h^0 t$ and $h^0 \to Z^0Z^0$ decays.}
\label{fig:Higgs-diagram}
\end{figure}

For example, with $M_h$ in the 200 - 230 GeV range, the standard model predicts a 
branching fraction of approximately 2.8\% for $h^0 \to Z^0Z^0$ with one $Z^0$ 
decaying to $e^+e^-$ or $\mu^+\mu^-$ and the other $Z^0$ decaying to jets.
Taking into account that $B(t^\prime \to h^0 t) \approx 20\%$ for $m_{t^\prime} = 450$ GeV,
and that $\sigma (p\bar{p} \to G_\mu^\prime \to t^\prime t)$ can be as large as 200 fb 
for $M_G = 1$ TeV (see Figure~\ref{fig:r_vs_sL}), 
we find that the process shown in Figure~\ref{fig:Higgs-diagram} can have a cross section 
of about 1 fb at the Tevatron, which is more than an order of magnitude larger than the 
corresponding $t\bar{t} h$ cross section in the Standard Model.
Thus, the CDF and D0 experiments
could each observe several $t\bar{t}h^0$ events with $h^0 \to Z^0Z^0 \to (\ell^+\ell^-)(jj)$. 
The background 
for this process is likely to be small, given that the two $Z$'s form a mass
peak at  $M_h$.  

Single-$t^\prime$ production from $W$ exchange is rather small at the 
Tevatron. Compared to the standard model single-$t$ production \cite{Willenbrock:1986cr},
single-$t^\prime$ production is suppressed due to the larger $t^\prime$ mass,
and also by $s_L^2$ arising from the $t^\prime$-$W$-$b$ vertex given in Eq.~(\ref{wtpb}).
We find $\sigma (p\bar{p} \to t^\prime j) \approx 4$ fb $(s_L/0.4)^2$ without cuts, 
for $m_{t^\prime} = 450$ GeV. It would be difficult to distinguish this signature 
from the large $W+2j$ background, which is estimated in Ref.~\cite{Atre:2008iu}.

At the LHC, by contrast, single-$t^\prime$ production is the main 
production mechanism for a heavy vectorlike $t^\prime$ \cite{Han:2005ru, Willenbrock:1986cr}.
The $t^\prime \rightarrow h^0 t$ decay followed by $h \to b\bar{b}$  has been 
studied in Ref.~\cite{Azuelos:2004dm}, while the more promising case 
$h^0\to Z^0 Z^0$ with one of the $Z^0$ bosons decaying leptonically
 has been discussed in  Ref.~\cite{:2006jm}.
The LHC signals arising from QCD $t^\prime\bar{t^\prime}$ production are discussed in 
\cite{Azuelos:2004dm, AguilarSaavedra:2005pv, Holdom:2007nw}.

The LHC phenomenology in the presence of both the $t^\prime$ and the 
gluon-prime has not been studied  yet.
The small parton distribution for antiquarks within the proton 
reduces the gluon-prime production in $pp$ collisions compared to $p\bar{p}$ collisions
(gauge invariance forbids the gluons to couple at tree level to a single $G^\prime_\mu$).
Nevertheless, the large center-of-mass energy planned for the LHC will significantly extend
the reach in $m_{t^\prime}$ and $M_G$.

\section{Conclusions}

We have shown that the pair production of top-prime 
quarks at the Tevatron may be increased by up to 
two orders of magnitude compared to the QCD process if it occurs via an 
$s$-channel resonance.  In the case where this resonance is a gluon-prime 
particle, the Tevatron searches may be sensitive to top-primes as heavy as 
600 GeV.  
Such a gluon-prime arises in gauge extensions of QCD, 
while the top-prime can arise from a vectorlike quark that has a
mass mixing with the standard top quark.

It is intriguing that the simple renormalizable model for new physics presented here, 
which includes
only two particles beyond the standard model, fits the noticeable CDF excess
in the semileptonic $(Wj)(Wj)$ channel without fine-tuning.
Fixing the top-prime mass at 450 GeV, we can match the 
CDF excess for any gluon-prime mass in the 850 - 1200 GeV range. 
A $Z^\prime \to t^\prime \bar{t^\prime}$ interpretation of this excess 
would be problematic, because the 
large coupling to $t^\prime$ would require the embedding of the associated gauge symmetry 
in a non-Abelian gauge group at a scale near the $Z^\prime$ mass.

Independent of the CDF excess, our $t^\prime + G_\mu^\prime$ model motivates 
new ways of analyzing existing Tevatron data.
Single production of a $t^\prime$ in association with a top quark may have a large cross section.
This process, as well as   $t^\prime \bar{t^\prime}$ 
production, with a top-prime subsequently decaying 
to $h^0 t$ imply that the Tevatron may be sensitive to Higgs masses
as large as 230 GeV. A possible discovery channel is resonant  
$t^\prime \bar{t}$ production 
followed by $t^\prime \to h^0 t$ and $h^0 \to Z^0Z^0$ decays.

On a more general note, the  $t^\prime + G_\mu^\prime$ model presented here underscores 
the importance of reconstructing the invariant mass of the entire event, 
even for complicated final states such as $WjWj$ or $t\bar{t}h^0$.
Furthermore, the rich phenomenology predicted in this simple extension of the 
standard model (including, for example, the displaced $t^\prime$ decays 
mentioned in Section 2, or the $WbtZ$ signature mentioned in Section 4) 
illustrates that the Tevatron may discover new physics
provided that searches are extended to cover as many signatures as possible.

\bigskip
{\bf Note added:} Subsequent to the completion of this paper, the D0 search for resonances in the $WbWb$ final state was updated with 3.6 fb$^{-1}$ of data \cite{D0-5882}.  
The new result includes 17 events in the 800 - 1000 GeV range, with a background 
of approximately 6.4 events. The extra events may arise from a $G^\prime_\mu$
resonance with $M_G \approx 900$ GeV which decays into a $t^\prime$ pair or
into a $t^\prime$ and a $t$ quark. A set of parameters consistent with 
this scenario is $m_{t^\prime} \approx 430$ GeV, $r \approx 0.2$, $s_L \approx 0.1$, 
$M_h \approx 120$ GeV.

\bigskip
{\bf Acknowledgments:} We would like to thank 
Anupama Atre, Gustavo Burdman, DooKee Cho, John Conway, Robin Erbacher,
Patrick Fox, Chris Hill, Andrew Ivanov, 
Christian Schwanenberger, Thomas Schwarz and Tim Tait 
for useful comments. 
Fermilab is operated by Fermi Research Alliance, LLC, under Contract
DE-AC02-07CH11359 with the US Department of Energy.


\vfil
\end{document}